\documentclass[iop]{emulateapj}
\usepackage{graphicx,amsmath,amssymb,natbib}
\shorttitle{FRB afterglow association}
\shortauthors{Vedantham et al.}

\begin{document}

\title{On associating fast radio bursts with afterglows}

\author{H.~K.~Vedantham\altaffilmark{1}, V.~Ravi\altaffilmark{1}, K.~Mooley\altaffilmark{2,4}, D.~Frail\altaffilmark{3}, G.~Hallinan\altaffilmark{1}, and S.~R.~Kulkarni\altaffilmark{1}}
\affil{$^{1}$Cahill Center for Astronomy and Astrophysics, MC 249-17, California Institute of Technology, Pasadena, CA 91125, USA; harish@astro.caltech.edu}
\affil{$^{2}$Oxford Centre for Astrophysical Surveys, University of Oxford, Denys Wilkinson Building, Keble Road, Oxford OX1 3RH}
\affil{$^{3}$National Radio Astronomy Observatory, 1003 Lopezville Road, Socorro, NM 87801-0387, USA}
\altaffiltext{4}{Hintze Fellow}

\begin{abstract}

A radio source that faded over six days, with a redshift of $z\approx0.5$ host, has been identified by \citet{keane2016} as the transient afterglow to a fast radio burst (FRB\,150418). We report follow-up radio and optical observations of the afterglow candidate and find a source that is consistent with an active galactic nucleus. If the afterglow candidate is nonetheless a prototypical FRB afterglow, existing slow-transient surveys limit the fraction of FRBs that produce afterglows to 0.25 for afterglows with fractional variation, $m=2|S_1-S_2|/(S_1+S_2)\geq0.7$, and 0.07 for $m\geq1$, at 95\% confidence. In anticipation of a barrage of bursts expected from future FRB surveys, we provide a simple framework for statistical association of FRBs with afterglows. Our framework properly accounts for statistical uncertainties, and ensures consistency with limits set by slow-transient surveys.  
%Afterglow associations with the barrage of bursts expected from future FRB surveys must satisfy constraints on the afterglow rate set by state of the art slow-transient surveys.

\end{abstract}

\keywords{methods: statistical -- astrometry -- radio continuum: galaxies}

\section{Introduction}  \label{sec:instr}

Fast radio bursts (FRBs) are millisecond-duration, intense ($\sim 1$~Jy) GHz transients that have dispersion measures that are well in excess of expected Milky Way contributions \citep{lorimer2007,thornton2013,spitler14,bb14,ravi15,petroff15,masui2015,keane2016}. Extragalactic FRBs would represent a truly extraordinary class of radio emitter \citep[for e.g.,][]{kashimaya2013,lyubarsky2014,kulkarni2014,cordes2016}. If FRBs originate at cosmological distances, studies of FRB samples will revolutionize our understanding of the intergalactic medium \citep[e.g.,][]{mcq14, zheng2014}. \\

Localization of an FRB to a host galaxy will not only determine the distance scale of FRBs, but will also provide vital clues regarding their origins, and realize the anticipated diagnostic of the IGM. \citet[][hereafter K16]{keane2016} promptly followed-up a Parkes event, FRB 150418. The field was imaged using the Australia Telescope Compact Array (ATCA) in the 4.5--8.5\,GHz band. The first observations began 2\,hr post-burst.  The subsequent four epochs were at 5.8, 7.8, 56, and 190\, days post-burst. Two variable sources were identified: a potential gigahertz-peaked spectrum source, and one that faded by a factor of $\sim$2.5 by the third epoch (7.8\,days).\footnote{The 5.5\,GHz flux-densities reported by K16 at the five epochs are $270\pm50$, $230\pm20$, $90\pm20$, $110\pm20$ and $90\pm20\,\mu$Jy respectively.}\\

The latter source, identified with a redshift of $z\approx 0.5$ galaxy, was interpreted by K16 to be the transient afterglow of FRB\,150418. To clearly distinguish this event from hypothetical FRB afterglows, we will refer to it as K16flare. K16 used previous surveys for week-timescale variables and transients \citep[e.g.,][]{bell2015,mooley2016} to determine a false alarm probability of $<0.1\%$ of observing K16flare in their observations. In addition, K16 interpreted the light curve of K16flare as being consistent with the radio emission sometimes observed following a short gamma-ray burst \citep{fong2015}. \\

The association between FRB~150418 and K16flare, if true, would be a spectacular confirmation of the cosmological nature of FRBs, enabling their application to intergalactic medium studies. However, even before the publication ink was dry, \citet{wb16} reported persistent radio emission from the host galaxy of K16flare 11 months after the FRB, brighter than the final K16 measurement, and thus suggested that it was an example of common  variability in Active Galactic Nuclei (AGNs) and was unrelated to  the FRB. Given the potential importance of K16's discovery, we consider the matter worthy of closer investigation.\\ 

The paper is organized as follows. In \S2, we present follow-up observations of the candidate FRB host galaxy with the Karl G. Jansky Very Large Array (JVLA), and the W. M. Keck Observatory. In \S3 we explore the hypothesis that K16flare is an AGN unrelated to the FRB. In \S4, we explore the consequences of  the K16flare-FRB association as asserted by K16. We present the implications of our study to future FRB afterglow searches in \S5, and
conclude with a summary in \S6.

\section{Radio and Optical Observations}\label{sec:obs}
\label{sec:RadioOptical}

\begin{figure*}
\centering
\includegraphics[width=0.75\linewidth]{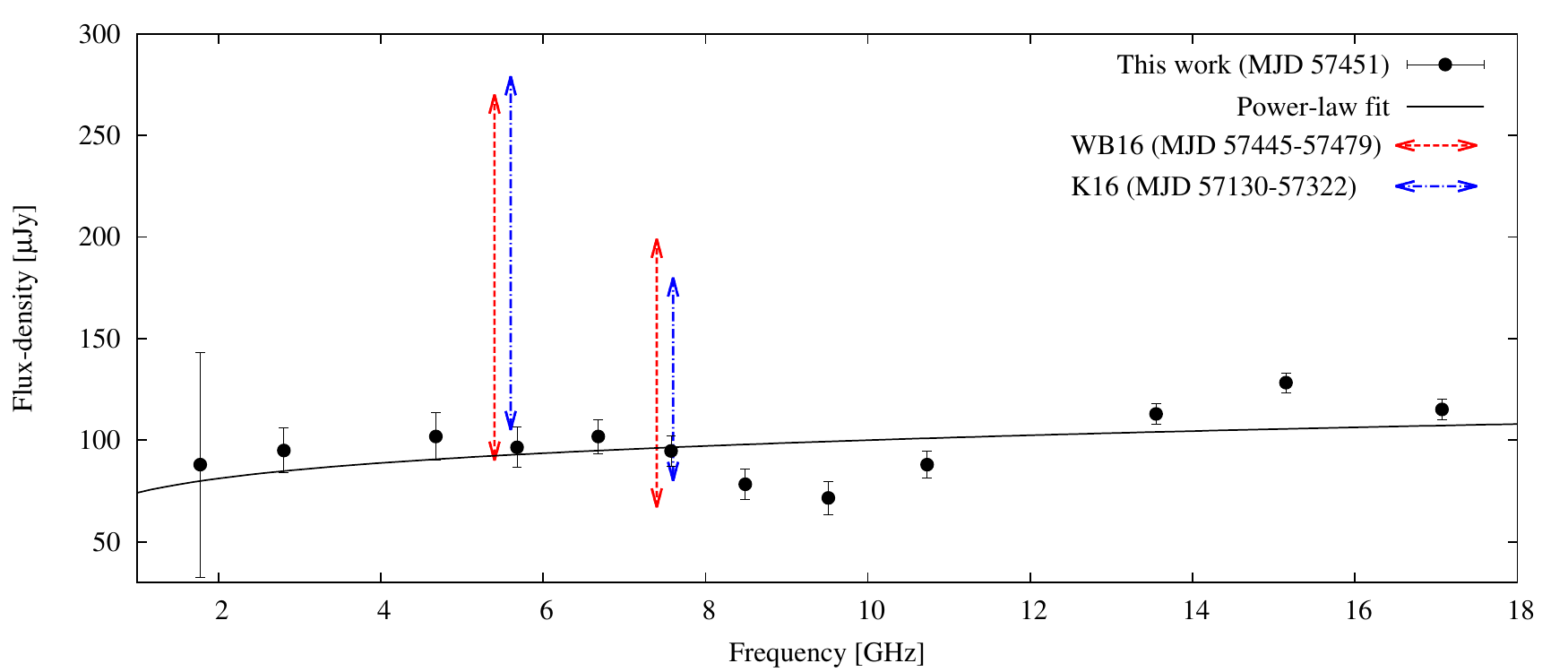}
\caption{JVLA radio spectrum of the host galaxy of K16flare, WISE\,J071634.59$-$190039.2 (black circles), obtained on MJD 57451. The solid black line shows a best-fit power-law spectrum to our data:
$S_\nu = (100\pm5)\left[\nu/10~{\rm GHz}\right]^\alpha$ with $\alpha=0.13\pm0.10$.
Also shown as vertical lines are the range of temporal variations seen at 5.5\,GHz and 7.5\,GHz by K16 (blue dot--dash) and subsequent variability (red dashed) reported by WB16 \citep{wb16,wbc16}. 
\label{spec}}
\end{figure*}

On 2016 March 04 (MJD 57451) we undertook observations over the frequency range 1--18\,GHz
of K16flare with the JVLA (DDT program 16A-432). Our observations were conducted during a 
single 3.5\,hr block. The JVLA was in the C configuration. We used standard wide-band continuum observing set-ups and 3C\,147 to place our observations on the Perley--Butler flux-density scale \citep{pb13}. The data were processed in CASA~4.5.2 with the standard NRAO pipeline.\footnote{https://science.nrao.edu/facilities/vla/data-processing/pipeline.} In the $L$-band (1.4\,GHz) the image rms was $50\,\mu$Jy whereas it ranged from 4 to 10$\,\mu$Jy across the  $S$--$Ku$ (2\,GHz$-$18\,GHz) band. We detected a point-like source across the entire decimetric band (Figure~\ref{spec}). The best-fit ($Ku$-band) position (J2000) is 07$^{\rm h}$16$^{\rm m}$34$^{\rm s}$.559(3), $-$19$^{\circ}$00$^{\rm m}$39$^{\rm s}$.73(7) ($1\sigma$ errors in the final significant figures in parentheses), which is consistent with that of K16flare.\\

\begin{figure}[hb]
\centering
\includegraphics[width=0.7\linewidth,angle=-90]{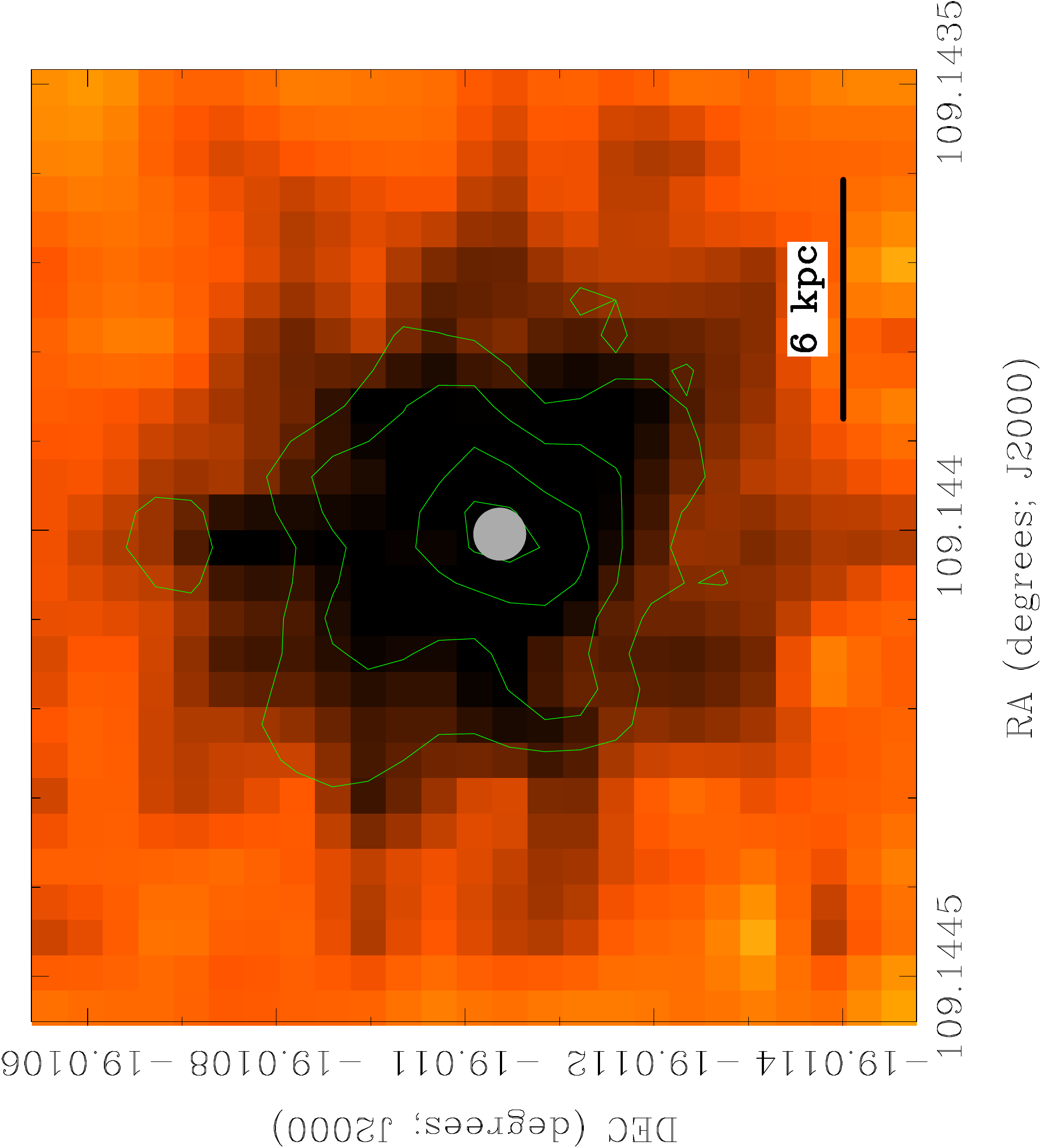}
\caption{Overlay of the $Ku$-band radio position of the source 
(grey circle, with 0.1\arcsec~radius; see \S\ref{sec:RadioOptical})
on a Ks-band image of WISE\,J071634.59$-$190039.2 which in turn was tied to the LRIS $R$-band image. 
The contours refer to the LRIS $R$-band image (levels:  $[3,5,7, 9]\sigma$). 
The scale bar corresponds to 6\,kpc at a redshift of 0.492, assuming
cosmological parameters measured by the {\it Planck} mission. 
\label{optical}}
\end{figure}

Separately, on MJD 57453, we observed the putative host galaxy (WISE\,J071634.59$-$190039.2)  with the Low Resolution Imaging Spectrograph \citep[LRIS;][]{lris} mounted on the Keck I telescope. We obtained three exposures in the g and R optical bands with Keck-1/LRIS in imaging mode, totaling 610\,s. Observing conditions were good, with 0\arcsec.75~$R$-band seeing. The data were initially processed using D. Perley's {\tt lpipe} software.\footnote{http://www.astro.caltech.edu/$\sim$dperley/programs} 
Using an initial 10\,s exposure, we obtained an initial astrometric solution from the USNO-B2 catalog using D. Perley's {\tt autoastrometry.py} software, and refined the astrometry using stars with Ks magnitudes between 10$-$14 from the 2MASS Point Source Catalog \citep[PSC;][]{scs06}. The PSC astrometric accuracy is 70--80\,mas: we assume a 0.1\arcsec~($1\sigma$) astrometric accuracy to account for possible minor distortion in the image. We then corrected the astrometry of our two 300\,s exposures using the shallow exposure, and co-added  the images. Separately, we obtained a deep Ks-band image of the field observed by M.\ Kasliwal (and presented
in K16). An overlay of the radio position on the final $R$-band and $K$-band images is shown in Figure~\ref{optical}.

\begin{figure*}
\centering
\includegraphics[width=0.75\linewidth]{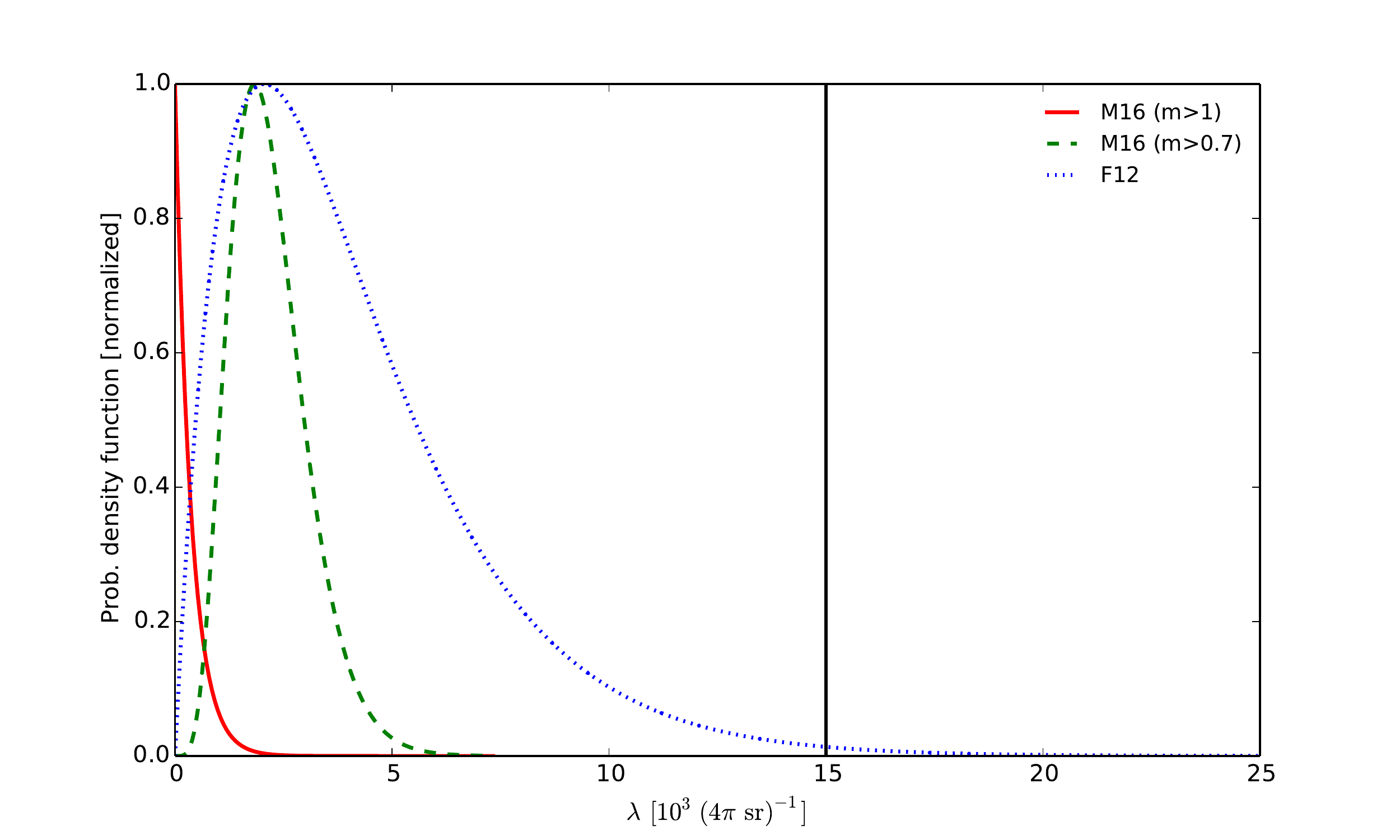}
\caption{Posterior probability density function for the areal density of afterglows, $\lambda$, from slow-transient surveys. The completeness limit is 270$\,\mu$Jy at 5.5\,GHz. The black vertical line represents the all-sky afterglow rate for an FRB rate of 2500\,sky$^{-1}$\,day$^{-1}$, and an afterglow duration of six days. K16flare has an $m$-value of $1\pm0.3$.\label{fig:poiss_post}}
\end{figure*}

\section{K16flare as a variable AGN} \label{sec:agn_interp}

\citet{wb16} note that the radio luminosity measured by K16, and the near-infrared colors of the host galaxy WISE\,J071634.59$-$190039.2, are consistent with that of a low-luminosity AGN. We note that the radio source continues to vary even a year after the FRB. \citet{wbc16} 
reported a flux-density of $157\pm6\mu$Jy (5~GHz band; 2016 Feb 27/28). Our  observations taken 
only six days later find the source to have decayed to 96$\pm$8\,$\mu$Jy. The fractional variation\footnote{We use the definition: $m=2|S_1-S_2|/(S_1+S_2)$, where $S_1$ and $S_2$ are the flux-densities at the two epochs.} between the two runs is $m=0.5\pm0.1$. For comparison, the maximum two-epoch fractional variation of K16flare in the K16 observations was $m=1.0\pm0.3$. Variability of $m\lesssim 1$ has been seen in other AGNs \citep{mooley2016}. \\

We measure the radio luminosity of the putative host to be $L\sim 10^{30}$\,erg\,s$^{-1}$\,Hz$^{-1}$. As shown by \citet{hbwv08}, it is not uncommon for elliptical galaxies without an optical signature of nuclear activity to harbor a low radio luminosity ($L\lesssim 10^{30}$\,erg\,s$^{-1}$\,Hz$^{-1}$) AGN. Furthermore, the spectrum seen in Figure~\ref{spec} is flat across the entire band (1--18\,GHz) and is consistent with that seen in several known AGN samples \citep{herbig1992,kkn+02}. The spectral bumps are suggestive of multiple, compact, optically thick synchrotron components. Variable radio emission from AGN cores in the  $\gtrsim 5$\,GHz band is not unusual, and typically originates in relativistic shocks of compact jets at milli-arcsecond scales \citep{bch+15}.\\

 Alternatively, for a brightness temperature of $T_{\rm B}=10^{12}$\,K that is typical of AGN cores, the angular size of the radio source is about $3\,\mu$as in size, which is comparable to the Fresnel scale for Galactic interstellar scattering. Refractive interstellar scintillations are common in this regime, with variations of $m\sim (\nu/\nu_0)^{17/30}$, on timescales of $\tau\sim 2(\nu_0/\nu)^{11/5}$, with a broad bandwidth of $\Delta\nu/\nu\sim 1$ \citep{walker1998}. Here, $\nu_0$ is the transition frequency below which we expect interstellar scattering to be strong. \citep{walker1998} estimate $\nu_0\approx 20\,$GHz for the low Galactic latitude ($b=3.2$\,deg) of K16flare, yielding $m\sim 0.5$, and $\tau\sim 2$\,days. Thus, both the variability seen in K16flare and subsequent observations, and the smoothly undulating spectrum presented here, are also consistent with refractive interstellar scintillation (see also \citet{aki2016} for more detailed arguments.) of a source with little or no intrinsic variability. Finally, as can be seen from Figure~\ref{optical}, the radio source coincides with the light centroid of the putative host galaxy to within experimental errors, $\lesssim0.1$\arcsec.\\

We thus conclude that the simplest hypothesis explaining (i) the persistence of a $\approx 100\,\mu$Jy flat-spectrum source nearly a year after the K16 observation of K16flare, (ii) its continued variability on 6 day timescales, and (iii) the nuclear origin, is that K16flare was an example of AGN variability (intrinsic and/or extrinsic) and is unrelated to the FRB. Despite this apparently compelling conclusion, in the next section, we explore observational constraints on possible FRB afterglows from existing radio surveys for transients and variables. 

\section{ K16flare as the FRB 150418 afterglow} \label{sec:frb_interp}
In the absence of any additional insight, we assume that K16flare is a prototypical FRB afterglow ($S\approx 270\,\mu$Jy at 5.5\,GHz, spectral index of $-0.7$ at maximum), and search for evidence of such afterglows in the VLA radio variability surveys of \citet[][hereafter M16]{mooley2016} and \citet[][hereafter F12]{frail2012}.
Details of these two surveys can be found in the Appendix. We adopt a conservative 
all-sky FRB rate of $\lambda_{\rm FRB}=2500/(4\pi)\,{\rm sr}^{-1}\,{\rm day}^{-1}$ for fluence $\mathcal{F}>2$\,Jy\,ms \citep{keane2015}.\\

Since each FRB afterglow lasts six days, in a survey whose cadence exceeds 6\,days, the expected slow-transient rate from FRB afterglows is $\lambda_{\rm AG}=6\lambda_{\rm FRB} = 0.364$\,deg$^{-2}$\,epoch$^{-1}$. 
If all FRBs have K16flare-like radio afterglows, the 50-square degree 3-epoch\footnote{The M16 survey had 4 epochs. We only consider the first 3 epochs here since the last epoch provided baselines of about 1 year and may contain examples of long term variability that is inconsistent with K16flare.} 3\,GHz survey of M16 
should have yielded about fifty five afterglows. They found none with $m\geq1$, and five with $m\geq0.7$. Next consider the the 944-epoch, 0.0225\,deg$^2$ slow-transient 5-GHz survey 
analyzed by F12. F12 should have seen eight
afterglows; they found just one. Clearly, existing slow-transient surveys show that only a small fraction of FRBs can generate K16flare-like afterglows.\\

To place limits on the fraction of FRBs that can generate afterglows, in Figure~\ref{fig:poiss_post}, we display the posterior probability density functions of the areal density of radio sources that vary on timescales of a week. The black vertical line shows the expected areal density of FRB afterglows assuming that all FRBs generate 6-day afterglows similar to K16flare. Even if FRBs are the only channel to create 6-day timescale transients, the slow-transient surveys limit the fraction of FRBs that produce ($S\geq270\,\mu$Jy) afterglows to $<0.25$, for $m\geq0.7$, and $<0.07$ for $m\geq1.0$ with 95\% confidence. Therefore, if FRBs produce afterglows, based on the measured average slow-transient rate, K16 had a $\lesssim 10\%$ chance of seeing an afterglow to FRB\,150418.
\\

\section{Guidance for Future FRB Afterglow Searches} \label{sec:discussion}

Our experience with FRB\,150418 (K16flare) has informed us of the potential pitfalls in associating FRBs with afterglows, particularly given the high all-sky FRB rate. The areal density of 6-day FRB afterglows at any given epoch ranges from  $\lambda_{\rm AG}= 0$ (FRBs are not associated with afterglows) to $\lambda_{\rm AG}=0.37\,{\rm deg}^{-2}$ (all FRBs are associated with afterglows). Depending on the fraction $f$ of FRBs associated with afterglows, FRB afterglows can therefore form an insignificant part of the transient sky, or completely dominate it. Hence, blind slow-transient surveys cannot be used to simply set a non-FRB related background false positive rate for afterglow discovery. Below, we outline a self-consistent approach for statistically relating FRBs with afterglow candidates.\\

Let slow-transient surveys yield a transient background rate (FRB related or otherwise) of $\lambda_{\rm BG}$\,deg$^{-2}$\,epoch$^{-1}$, and let FRBs be localized to within $\Omega$~deg$^2$. We wish to determine the fraction $f$ of FRBs that yield afterglows. The detection of $n$ afterglow candidates in $N$ FRB follow-ups will yield the estimate: $f=n/N-\lambda_{\rm BG}\Omega$.  Based on Poisson statistics, the $1\sigma$ error on our estimate of $f$ will be $\approx \sqrt{n}/N$ for large $n$. For instance, detection of $n=100$ transients in follow-ups, will constrain $f$ with about 10\% fractional error ($1\sigma$).\\

New surveys such as the VLA Sky Survey \citep{vlass} will systematically explore the sub-mJy transient sky in the decimetric band and constrain the event background, $\lambda_{\rm BG}$. Coincidentally, upcoming FRB-machines such as CHIME \citep{chime} and UTMOST \citep{utmost} are expected to discover a barrage of FRBs ($\gtrsim1$\,day$^{-1}$). With the anticipated large FRB sample with prompt follow-up, the above framework will enable a direct measurement of the fraction $f$ of FRBs that are associated with transient afterglows.\\

We finally note that, while a statistical argument for FRB-afterglow association based on a large number of FRB follow-ups will be compelling, future localization of an FRB itself \citep[see ][]{law2015} at a few arcsecond-level would imply an (almost) absolute confirmation of the host galaxy.

\section{Summary}

We conducted radio and optical follow-up observations of the afterglow candidate to FRB~150418 (K16flare). We detected persistent radio emission from the host galaxy of K16flare $\sim 1$~year after the FRB, which is nuclear in origin (0.1\arcsec~astrometric precision), and has a flat radio spectrum (1--18\,GHz). It is therefore consistent with an AGN core, and does not present prima facie evidence of being associated with FRB\,150418.\\ 

If K16flare is nonetheless a prototypical FRB afterglow, existing slow radio transient surveys limit the fraction of FRBs that produce afterglows to $<0.25$ for fractional variation of $m\geq0.7$, and $<0.07$ for $m\geq1.0$ (95\% confidence). Finally, keeping upcoming FRB surveys in mind, we have presented a statistical framework to associate FRBs with afterglow candidates, which will determine the fraction of FRBs that produce afterglows.\\

\acknowledgments

H.K.V. thanks A.G.~de Bruyn and Anthony Readhead for insightful discussions. K.P.M. acknowledges 
support from the Hintze Foundation.  We thank Mansi Kasliwal and Jacob Jencsen for 
providing us with a deep K-band image of the field of K16flare.
The National Radio Astronomy Observatory is a facility of the National
Science Foundation operated under cooperative agreement
by Associated Universities, Inc. We thank NRAO for accepting our proposal (under the
Director's Discretionary program) for
rapid observations of K16flare's host galaxy.
Some of the data presented
herein were obtained at the W.M. Keck Observatory, which
is operated as a scientific partnership among the California
Institute of Technology, the University of California and the
National Aeronautics and Space Administration. The Observatory
was made possible by the generous financial support
of the W.M. Keck Foundation. 
%The authors wish to recognize and acknowledge the very significant cultural role and reverence that the summit of Mauna Kea has always had within the indigenous Hawaiian community.  We are most fortunate to have the opportunity to conduct observations from this mountain. 
%This publication makes use of data products from the Two Micron All Sky Survey, which is a joint project of the University of Massachusetts and the Infrared Processing and Analysis Center/California Institute of Technology, funded by the National Aeronautics and Space Administration and the National Science Foundation. 
This research has made use of ADS, CDS (Vizier and SIMBAD), NED, SDSS, IRSA and 2MASS.
{\it Facilities:} \facility{Jansky VLA}, \facility{Keck/LRIS}

\appendix

\section*{Limits from slow-transient surveys}

Afterglows probably emanate from expanding relativistic plasma where synchrotron self-absorption may be important. For this reason, apart from the M16 survey (2--4~GHz) which K16 consider in their false-positive rate calculation, we also consider limits on the transient areal density at $5$~GHz by \citet[][hereafter F12]{frail2012}. F12's survey is at a similar frequency as K16flare, and has undergone rigorous tests to rule out false-candidates due to imaging and interference-related artifacts\footnote{The \citet{frail2012} results were obtained by reprocessing a dataset original presented by \citet{bower2007}.}\\

The relevant survey parameters and findings are summarized in Table~\ref{tab:surveys}. The M16 survey has a completeness limit of $S=500\,\mu$Jy at 3\,GHz, or $327\,\mu$Jy at 5.5\,GHz assuming the same spectral index as that of K16flare. M16 found no transients, and no variables with $m\geq 1$. Though M16 list 10 variables (their Table~3) with $m\geq0.7$, half of them are grossly inconsistent with K16flare; their flux density drops and rises again on a 1~month timescale.
%M16 classify variables by their $m$-values: $m=2|S_1-S_2|/(S_1+S_2)$; $S_1$ and $S_2$ are flux-densities between epochs. For comparison, the $m$-value for K16flare is $m=1\pm0.3$. \\

The F12 survey had a completeness limit of $S=300\,\mu$Jy at 5\,GHz (280$\,\mu$Jy at 5.5\,GHz), and found just 1 transient; RT~19970528 was seen in their single-epoch search and faded from $1731\pm232\,\mu$Jy to $<37\,\mu$Jy within seven days. As such, it is similar to the K16 afterglow in its duration, but significantly brighter.\\

To compare the survey limits and the K16 afterglow on equal footing, we have: (i) computed a `5.5~GHz equivalent' completeness limit assuming a spectral index of $-0.7$, (ii) obtained the 95\% confidence limits on the Poisson parameter $\lambda$ in units of $(4\pi\,{\rm sr})^{-1}$, and finally (iii) scaled the limits to a completeness flux-density of $270\,\mu$Jy by assuming a uniformly distributed population in Euclidean space. The final limits on $\lambda$ are presented in the last column of Table \ref{tab:surveys} and in Figure~\ref{fig:poiss_post}. These limits on the slow-transient areal density are valid for \emph{any} afterglow, FRB-related or otherwise.

%\section*{A2: Afterglow rate calculation}

%Let $N_a$ be the number of FRBs expected within the slow-transient survey area per day. Let each FRB afterglow last $\tau_a$ days. According to K16, $\tau_a\approx 6$~days. Let the slow-transient survey cadence be $\tau_s$~days. The total numbers of FRB afterglows potentially detectable in a survey with $N_e$ epochs is then $N_{agl}=N_eN_a{\rm Min}(\tau_a,\tau_s)$ where Min(.) returns the smallest argument.\\ 

%Since for both the M16 and F12 surveys $\tau_s>\tau_a$, the expected \emph{all-sky} FRB afterglow rate is $N_{e}\tau_a\times$~(FRB all-sky rate). If all FRBs produce an afterglow similar to the one reported by K16, then for an FRB rate of $2500$\,sky$^{-1}$\,day$^{-1}$, the expected number of \emph{detectable} afterglows in the M16 and F12 surveys are 55 and 8 respectively.\\

\begin{table}
\centering
\caption{Parameters of the M16 and F12 slow-transient surveys. FoV is the field of view; $N_e$ is number of survey epochs; $N_{det}$ is the number of transients or variables detected; $m$ is the fractional variation; $\lambda$ is the Poisson rate parameter scaled to a completeness-limit of 270$\mu$Jy at 5.5~GHz. \label{tab:surveys}} 
\begin{tabular}{lllll}
\hline
Survey 	& FoV [deg$^{2}$]	& $N_e$	& $N_{det}$	& 95\% CL on $\lambda$\\
		&					&		&			&	[ ($4\pi$sr)$^{-1}$ ] \\\hline
        &	&	&	& \\
M16 ($m\geq1$)		& 52				& 3		& 0			& $0^{+1099}_{-0}$\\
M16 ($m\geq0.7$)  	& 52				& 3		& 5			& $1834^{+2021}_{-1111}$ \\
F12				& 0.0225			& 944	& 1		& $2058^{+7706}_{-1953}$ \\ \hline
\end{tabular}
\end{table}

\end{document}